\documentclass[journal,comsoc]{IEEEtran}
\usepackage[T1]{fontenc}
\usepackage{color}
\usepackage{amsmath}
\usepackage{url}
\usepackage{array}
\usepackage{fancyhdr}
\usepackage{latexsym}
\usepackage{diagbox}
\usepackage{epsfig}
\usepackage{amssymb}
\usepackage{amsfonts}
\usepackage{amsxtra}
\usepackage{xspace}
\usepackage{makeidx}
\usepackage{graphics}
\usepackage{theorem}
\usepackage{subfigure}
\usepackage{multirow}
\usepackage{verbatim}
\usepackage{algorithm}
\usepackage{algorithmic}
\DeclareMathOperator*{\minimize}{minimize}
\DeclareMathOperator*{\maximize}{maximize}
\usepackage[noadjust]{cite}

\newtheorem{proposition}{\textbf{Proposition}}

\begin{document}
\title{\huge{Joint Active and Passive Beamforming for Intelligent Reflecting Surface Aided Multiuser MIMO Communications}}

\author{Xingyu Zhao,~\IEEEmembership{Student Member,~IEEE,} Tian Lin,~\IEEEmembership{Student Member,~IEEE,} and Yu Zhu,~\IEEEmembership{Member,~IEEE}
\thanks{This work was supported by National Natural Science Foundation of China under Grant No. 61771147.}
\thanks{X. Zhao, T. Lin, and Y. Zhu are with the Department of Communication Science and Engineering, Fudan University, Shanghai, China (e-mail: xingyuzhao19@fudan.edu.cn, lint17@fudan.edu.cn, zhuyu@fudan.edu.cn).}}

\maketitle

\begin{abstract}

This letter investigates the joint active and passive beamforming optimization for intelligent reflecting surface (IRS) aided multiuser multiple-input multiple-output systems with the objective of maximizing the weighted sum-rate. We show that this problem can be solved via a matrix weighted mean square error minimization equivalence. In particular, for the optimization of the passive IRS beamforming, we first propose an iterative algorithm with excellent performance based on  manifold optimization. By using fractional programming to obtain a more tractable object function, we then propose a low complexity algorithm based on majorization-minimization. Numerical results verify the convergence of our proposed algorithms and the significant performance improvement over the communication scenario without IRS assistance.

\end{abstract}


\IEEEpeerreviewmaketitle

\section{Introduction}\label{sec:introduction}
Intelligent reflecting surface (IRS) has been recently recognized as an attractive technology for future wireless communications due to its advantage of re-configuring the radio signal propagation environment with limited power consumption \cite{Yu:2020,Huang:2019,Guo:2020}. Among the research problems for the IRS-aided systems, there have been many works focusing on the joint active and passive beamforming optimization with different objectives to improve the multiuser communication performance  \cite{Guo:2020,Li:2020,WU:2019}. For example, the authors in \cite{Guo:2020} studied the weighted sum rate (WSR) maximization problem by using fractional programming (FP) \cite{Shen:2019}. In \cite{Li:2020}, the authors investigated the same problem by using the Lagrangian method for the active beamforming design and manifold optimization (MO) for the IRS passive beamforming design. The authors in \cite{WU:2019} studied the total transmit power minimization problem with the constraint on each user's signal-to-interference-plus-noise ratio. However, all of these works focused on the multiuser multiple-input single-output scenario, and thus it is of interest to study the multiuser multiple-input multiple-output (MIMO) system with IRS assistance. 



In this letter, we aim at the joint optimization of the transceiver active beamforming and the IRS passive beamforming for IRS-aided multiuser MIMO systems with the objective of maximizing the WSR. We show that this problem can be solved via a matrix weighted mean square error (MW-MSE) minimization equivalence. In particular, the MW-MSE minimization problem can be decomposed into the transceiver beamforming and IRS passive beamforming optimization subproblems using alternating optimization (AO). To deal with the non-convex IRS beamforming optimization subproblem, we first propose an MO based iterative algorithm. We also propose a low complexity algorithm by adopting FP to obtain a more tractable problem and applying majorization-minimization (MM) \cite{Sun:2017} to solve it. We prove the convergence of these two algorithms in WSR. Numerical results show that our proposed algorithms achieve significant performance improvement over the communication scenario without IRS assistance.



\emph{Notations:} ${\mathbf{a}}\left( i \right)$ denotes the $i$-th element of vector $\mathbf{a}$. ${\left(  \cdot  \right)^ * }$, ${\left(  \cdot  \right)^T}$, and ${\left(  \cdot  \right)^H}$ respectively denote the complex conjugate, transpose, and complex conjugate transpose of a matrix (vector). ${\text{tr}}\left( \mathbf{A} \right)$ and ${\left\|  \mathbf{A}  \right\|_F}$ denotes the trace and Frobenius norm of matrix $\mathbf{A}$, respectively. $\odot$ is the matrix Hadamard product operator. $\mathrm{Re} \left\{  \cdot  \right\}$ and $\angle \left(  \cdot  \right)$ denote the real part and the angle of a complex number, respectively. ${\text{diag}}\left( {\mathbf{a}} \right)$ is a diagonal matrix with the entries of vector $\mathbf{a}$ on its main diagonal. ${\text{blkdiag}}({{\mathbf{A}}_1}, \ldots ,{{\mathbf{A}}_N})$ returns a block diagonal matrix with sub-matrices ${{\mathbf{A}}_1}, \ldots ,{{\mathbf{A}}_N}$. ${\lambda _{\max }}\left( {\mathbf{A}} \right)$ denotes the largest eigenvalue of matrix $\mathbf{A}$. ${\mathbf{I}}$ denotes the identity matrix.

\section{System Model}

Consider the downlink transmission of an IRS-aided multiuser MIMO system, where a base station (BS) equipped with ${N_{\text{t}}}$ antennas serves $K$ users, each of which is equipped with ${N_{\text{r}}}$ antennas, with the aid of an IRS implemented by $N$ configurable phase shifters. With the support of the beamforming at the transceiver and the reflection operation on the IRS, the resulting signal at the $k$-th user can be represented as
\begin{equation}
\label{receie singal}
{{\mathbf{y}}_k} = {\mathbf{W}}_k^H({{\mathbf{H}}_{{\text{r}},k}}{\mathbf{\Theta G}} + {{\mathbf{H}}_{{\text{d}},k}})\sum\limits_{i = 1}^K {{{\mathbf{F}}_i}{{\mathbf{s}}_i}}  + {\mathbf{W}}_k^H{{\mathbf{z}}_k},
\end{equation}
where ${{\mathbf{H}}_{{\text{d}},k}} \in {\mathbb{C}^{{N_{\text{r}}} \times {N_{\text{t}}}}}$, ${{\mathbf{H}}_{{\text{r}},k}} \in {\mathbb{C}^{{N_{\text{r}}} \times N}}$, and ${\mathbf{G}} \in {\mathbb{C}^{N \times {N_{\text{t}}}}}$ denote the channel matrices from the BS to the $k$-th user, from the IRS to the $k$-th user, and from the BS to the IRS, respectively. $\sum\nolimits_{i = 1}^K {{{\mathbf{F}}_i}{{\mathbf{s}}_i}}  = {\mathbf{Fs}}$ denotes the transmitted signal at the BS, where ${\mathbf{s}} = {\left[ {{\mathbf{s}}_1^T, \cdots ,{\mathbf{s}}_K^T} \right]^T} \in {\mathbb{C}^{K{N_{\text{r}}} \times {1}}}$ with $\mathbb{E}\left[ {{\mathbf{s}}{{\mathbf{s}}^H}} \right] = \mathbf{I}$ denotes the transmitted symbols for all users and is processed by the transmit beamformer ${\mathbf{F}} = \left[ {{{\mathbf{F}}_1}, \cdots ,{{\mathbf{F}}_K}} \right] \in {\mathbb{C}^{{N_{\text{t}}} \times K{N_{\text{r}}}}}$ with a maximum power ${P_{\text{t}}}$ constraint, i.e., $\sum\nolimits_{i = 1}^K {\left\| {{{\mathbf{F}}_k}} \right\|_F^2 \leq {P_{\text{t}}}}$. The passive beamforming matrix at the IRS is denoted by ${\mathbf{\Theta }} = {\text{diag}}( {{{[ {{e^{j{\theta _1}}}, \cdots ,{e^{j{\theta _N}}}} ]}^T}} )$, where ${\theta _i} \in \left[ {0,2\pi } \right]$ represents the phase shift of the $i$-th reflection element. ${{\mathbf{W}}_k} \in {\mathbb{C}^{{N_{\text{r}}} \times {N_{\text{r}}}}}$ denotes the receive beamformer of the $k$-th user. Finally, ${{\mathbf{z}}_k}$ represents the noise at the $k$-th user satisfying the circularly symmetric Gaussian distribution with zero mean and covariance matrix $\sigma^2\mathbf{I}$. Throughout this letter, we focus on the optimization of the transceiver beamformers and the IRS phase shifts to maximize the WSR and assume that perfect channel state information is available. It can be found that the WSR $R$ is given by 
\begin{equation}
\label{weight-sum-rate}
R = \sum\limits_{k = 1}^K {{\omega _k}{R_k}}  = \sum\limits_{k = 1}^K {{\omega _k}\log \left| {{\mathbf{I}} + {\mathbf{F}}_k^H{\mathbf{H}}_k^H{\mathbf{\Lambda }}_k^{ - 1}{{\mathbf{H}}_k}{{\mathbf{F}}_k}} \right|} , 
\end{equation}
where ${{\omega _k}}$ and  $R_k$ represent  the priority and rate (spectral efficiency) of the $k$-th user, respectively, 
${{\mathbf{H}}_k} = {{\mathbf{H}}_{{\text{r}},k}} {\mathbf{\Theta G}} + {{\mathbf{H}}_{{\text{d}},k}}$ denotes the combined channel response, and ${{\mathbf{\Lambda }}_k} = {\sigma ^2}{\mathbf{I}} + \sum\nolimits_{i \ne k}^K {{{\mathbf{H}}_k}{{\mathbf{F}}_i}{\mathbf{F}}_i^H{\mathbf{H}}_k^H}$ denotes the effective noise covariance matrix associated with the $k$-th user. Hence, the WSR maximization problem can be formulated as
\begin{equation}\label{prb:rate}
\begin{array}{cl}
\displaystyle{\maximize_{{\mathbf{F}},{\mathbf{\Theta }}}} & \sum\limits_{k = 1}^K {{\omega _k}{R_k}}   \\
{\text{s}}{\text{.t}}{\text{.}} & \left\| {\mathbf{F}} \right\|_F^2  \leq {P_{\text{t}}}\\ \quad & {\theta _i} \in \left[ {0,2\pi } \right],\quad \forall i = 1,\cdots, N.
\end{array}
\end{equation}

\section{Joint Transceiver and IRS Design}\label{sec:Design}

Aiming at solving the highly non-convex problem of \eqref{prb:rate}, in this section, we show that the original problem can be solved equivalently through an MW-MSE minimization problem, which is then decomposed into the transceiver beamforming and the IRS phase shifts optimization subproblems. We then propose two algorithms for the IRS phase shifts optimization with guaranteed convergence, and finally analyze the computational complexity. 

\subsection{MW-MSE Minimization Problem}\label{subsec:MW-MSE-idea}
It has been shown in \cite{Shi:2011,Christensen:2008,xingyu:2020} that the WSR maximization problem can be solved equivalently by solving an MW-MSE minimization problem in MIMO systems. This method can be directly applied to solve the problem in \eqref{prb:rate} as the IRS can be regarded as part of the channel between the BS and the users. Furthermore, similar to that in \cite{xingyu:2020,Lin:2019,Joham:2015}, to  facilitate handling the power constraint, an auxiliary scalar factor $\beta$ is introduced to represent the original transmit beamformer ${\mathbf{F}}$ as ${\mathbf{F}} = \beta {\mathbf{\tilde F}}$. By defining a positive semi-definite weighting matrix as ${{\mathbf{\Gamma }}_k} \succeq {\bf{0}}$ for each user, the MW-MSE minimization problem for equivalently solving (\ref{prb:rate}) is given by
\begin{equation}\label{prb:wmmse}
\begin{array}{cl}
\displaystyle{\minimize_{{{\mathbf{W}}_k},\beta, {{{\mathbf{\tilde F}}}},{\mathbf{\Theta }},{{{\mathbf{\Gamma }}_k}}}} & {f_1}({{\mathbf{W}}_k},\beta ,{\mathbf{\tilde F}},{\mathbf{\Theta }},{{\mathbf{\Gamma }}_k})  \\
{\text{s}}{\text{.t}}{\text{.}} & \left\| {{{\mathbf{\tilde F}}}} \right\|_F^2  \leq {\beta ^{ - 2}}{P_{\text{t}}}\\ \quad & {\theta _i} \in \left[ {0,2\pi } \right],\quad \forall i = 1, \cdots ,N,
\end{array}
\end{equation}
where 
\begin{equation}
\label{func-pro:wmmse}
    {f_1}({{\mathbf{W}}_k},\beta ,{\mathbf{\tilde F}},{\mathbf{\Theta }},{{\mathbf{\Gamma }}_k}) = \sum\limits_{k = 1}^K {\operatorname{tr} ({{\mathbf{\Gamma }}_k}{{\mathbf{E}}_k}) - {\omega _k}\log \left| {\omega _k^{ - 1}{{\mathbf{\Gamma }}_k}} \right|} 
\end{equation}
and the MSE matrix ${{{\mathbf{E}}_k}}$ is given by
\begin{equation}
\label{MSE_matrix}
\begin{aligned}
{{{\mathbf{E}}_k}} & = \mathbb{E}\left[ {({\beta ^{ - 1}}{{\mathbf{y}}_k} - {{\mathbf{s}}_k}){{({\beta ^{ - 1}}{{\mathbf{y}}_k} - {{\mathbf{s}}_k})}^H}} \right]\\& = {\mathbf{I}} + {\beta ^{ - 2}}{\mathbf{W}}_k^H{{\mathbf{H}}_k}{\mathbf{F}}{{\mathbf{F}}^H}{\mathbf{H}}_k^H{{\mathbf{W}}_k}+ {\beta ^{ - 2}}{\sigma ^2}{\mathbf{W}}_k^H{{\mathbf{W}}_k}
\\& \quad - 2{\beta ^{ - 1}}\operatorname{Re} \left\{ {{\mathbf{W}}_k^H{{\mathbf{H}}_k}{{\mathbf{F}}_k}} \right\}.
\end{aligned}
\end{equation}

Although the problem in \eqref{prb:wmmse} is still highly non-convex, it can be decomposed into the transceiver beamforming and the IRS phase shifts optimization subproblems. 

\subsection{Transceiver Beamforming Optimization}\label{subsec:TR-BF}
For a fixed ${\mathbf{\Theta }}$, the problem in \eqref{prb:wmmse} reduces to the conventional transceiver beamforming optimization and has been extensively studied in \cite{Shi:2011} and \cite{Christensen:2008}. In particular, there exists a closed-form solution of the optimal receive beamformer of the $k$-th user by differentiating $f_1$ in \eqref{prb:wmmse} respect to $\mathbf{W}_k^*$ and setting the result to zero. That is, 
\begin{equation}
\label{beamforming-receiver}
{\mathbf{W}}_k = \beta {({{\mathbf{\Lambda }}_k} + {{\mathbf{H}}_k}{{\mathbf{F}}_k}{\mathbf{F}}_k^H{\mathbf{H}}_k^H)^{ - 1}}{{\mathbf{H}}_k}{{\mathbf{F}}_k}.
\end{equation}
 
For the transmit beamforming optimization, according to the method in \cite{Joham:2015} and with the  Karush-Kuhn-Tucker (KKT) conditions,  the optimal ${{\mathbf{\tilde F}}}$ and $\beta$ are given by  
\begin{equation}
\label{eqn:F}
{\mathbf{\tilde F}} = {\left( {{{\mathbf{H}}^H}{\mathbf{W\Gamma }}{{\mathbf{W}}^H}{\mathbf{H}} + \varphi {\mathbf{I}}} \right)^{ - 1}}{{\mathbf{H}}^H}{\mathbf{W\Gamma }}, \quad \beta = \sqrt{\frac{P_\mathrm{t}}{\|\mathbf{\tilde F} \|_F^2}}, 
\end{equation}
where ${\mathbf{W}}={\text{blkdiag(}}{{\mathbf{W}}_1}{\text{,}} \cdots {\text{,}}{{\mathbf{W}}_K}{\text{)}}$, ${\mathbf{\Gamma }} = {\text{blkdiag(}}{{\mathbf{\Gamma }}_1}{\text{,}} \cdots {\text{,}}{{\mathbf{\Gamma }}_K}{\text{)}}$, ${{\varphi  = {\sigma ^2}{\text{tr(}}{\mathbf{\Gamma }}{{\mathbf{W}}^H}{\mathbf{W}}{\text{)}}} \mathord{\left/
 {\vphantom {{\varphi  = {\sigma ^2}{\text{tr(}}{\mathbf{\Gamma }}{{\mathbf{W}}^H}{\mathbf{W}}{\text{)}}} {{P_{\text{t}}}}}} \right.
 \kern-\nulldelimiterspace} {{P_{\text{t}}}}}$, and 
\begin{equation*}
   {\mathbf{H}} = \underbrace {{{\left[ {{\mathbf{H}}_{{\text{r,1}}}^T, \cdots ,{\mathbf{H}}_{{\text{r,}}K}^T} \right]}^T}}_{{{\mathbf{H}}_{\text{r}}}}\mathbf{\Theta} {\mathbf{G}} + \underbrace {{{\left[ {{\mathbf{H}}_{{\text{d,1}}}^T, \cdots ,{\mathbf{H}}_{{\text{d,}}K}^T} \right]}^T}}_{{{\mathbf{H}}_{\text{d}}}} = {{\mathbf{H}}_{\text{r}}}\mathbf{\Theta} {\mathbf{G}} + {{\mathbf{H}}_{\text{d}}}.
\end{equation*}

Finally, the optimal weighting matrix can be obtained by differentiating $f_1$ with respect to ${{\mathbf{\Gamma }}_k}$ and setting the result to zero, which has a closed-form solution as follows
\begin{equation}\label{weighted-matrix}
     \boldsymbol{\Gamma}_{k}=\omega _k \mathbf{E}_{k}^{-1}.
\end{equation}

\subsection{IRS Phase Shifts Optimization}\label{subsec:IRS}
By substituting the above optimized transmitter beamformer into the problem in (\ref{prb:wmmse}) and ignoring other irrelevant constant items, the subproblem of IRS phase shifts optimization can be represented as
\begin{equation}\label{prb:irs}
\begin{array}{cl}
\displaystyle{\minimize_{{\mathbf{\Theta }}}} & {f_2}({\mathbf{\Theta }}) = \operatorname{tr} ({({{\mathbf{\Gamma }}^{ - 1}} + {\varphi ^{ - 1}}{{\mathbf{W}}^H}{\mathbf{H}}{{\mathbf{H}}^H}{\mathbf{W}})^{ - 1}})  \\
{\text{s}}{\text{.t}}{\text{.}}& {\theta _i} \in \left[ {0,2\pi } \right],\quad \forall i = 1, \cdots ,N.
\end{array}
\end{equation}

Nevertheless, it is intractable to deal with (\ref{prb:irs}) as each element of the IRS only causes a phase shift to the incident signal without affecting the amplitude, which results in a highly non-convex unit modulus constraint on each IRS element. In some previous works (e.g., \cite{Lin:2019} and \cite{Yu:2016}), MO has been shown to be an effective method to deal with such non-convex constraint with excellent performance. However, the application of MO is not straightforward and highly depends on the specific optimization problem. Furthermore, the MO-based algorithm normally requires high computational complexity due to the calculation of conjugate gradients containing the matrix inversion operation and the calculation of the optimal step size in the conjugate gradient algorithm. Thus, it is also significant to seek a lower complexity algorithm. In the following, we first propose a MO-based IRS phase shifts optimization algorithm and then a low complexity algorithm based on MM. 

 
\subsubsection{MO-IRS Algorithm}\label{subsubsec:MO}
To apply MO, the most crucial and difficult step is to derive the Euclidean conjugate gradient with respect to the diagonal matrix $\mathbf{\Theta}$, which is denoted by $\nabla {f_2}({\mathbf{\Theta }})$. Defining ${\mathbf{M}} = {{\mathbf{\Gamma }}^{ - 1}} + {\varphi ^{ - 1}}{{\mathbf{W}}^H}{\mathbf{H}}{{\mathbf{H}}^H}{\mathbf{W}}$ in \eqref{prb:irs}, the differential of $\text{d}(f_2(\mathbf{\Theta}))$ without considering the specific matrix structure of $\mathbf{\Theta}$ can be first derived as 
\begin{equation}
    \begin{aligned}
    {\text{d(}}{f_2}({\mathbf{\Theta }}){\text{)}}&\stackrel{(a)}= - \operatorname{tr} ({{\mathbf{M}}^{ - 2}}{\text{d}}({\mathbf{M}}))\\
    &\stackrel{(b)}= - {\text{tr(}}{{\mathbf{M}}^{ - 2}}({\varphi ^{ - 1}}{{\mathbf{W}}^H}{{\mathbf{H}}_{\text{r}}}\Theta {\mathbf{G}}{{\mathbf{G}}^H}{\text{d}}({{\mathbf{\Theta }}^H}){\mathbf{H}}_{\text{r}}^H{\mathbf{W}}\\& \quad  + {\varphi ^{ - 1}}{{\mathbf{W}}^H}{{\mathbf{H}}_{\text{d}}}{{\mathbf{G}}^H}{\text{d}}({{\mathbf{\Theta }}^H}){\mathbf{H}}_{\text{r}}^H{\mathbf{W}}))\\& \stackrel{(c)}=  - {\varphi ^{ - 1}}{\text{tr}}(({\mathbf{A}} + {\mathbf{B}}){\text{d}}({{\mathbf{\Theta }}^H})),
    \end{aligned}
\end{equation}
where ($a$) and ($b$) follow from the properties of matrix differentiation, and ($c$) follows from the properties of trace \cite{Zhangxianda} and the definitions of ${\mathbf{A}}  = {\mathbf{H}}_{\text{r}}^H{\mathbf{W}}{{\mathbf{M}}^{ - 2}}{{\mathbf{W}}^H}{{\mathbf{H}}_{\text{r}}}\mathbf{\Theta} {\mathbf{G}}{{\mathbf{G}}^H}$ and ${\mathbf{B}}  = {\mathbf{H}}_{\text{r}}^H{\mathbf{W}}{{\mathbf{M}}^{ - 2}}{{\mathbf{W}}^H}{{\mathbf{H}}_{\text{d}}}{{\mathbf{G}}^H}$. By using the important property of the matrix differentiation that ${\text{d(}}{f_2}({\mathbf{\Theta }}){\text{)}} = {\text{tr}}(\frac{{\partial {f_2}({\mathbf{\Theta }})}}{{\partial {{\mathbf{\Theta }}^ * }}}{\text{d(}}{{\mathbf{\Theta }}^H}{\text{)}})$ \cite{Zhangxianda}, and following the method in \cite{xingyu:2020} to derive the Euclidean conjugate gradient with respect to a matrix having a specific structure, we finally have 
\begin{equation}
\label{eqn:gra-theta}
\nabla {f_2}({\mathbf{\Theta }}){\text{   =  }}\frac{{\partial {f_2}({\mathbf{\Theta }})}}{{\partial {{\mathbf{\Theta }}^ * }}} \odot {\mathbf{I}} =  - {\varphi ^{ - 1}}({\mathbf{A}} + {\mathbf{B}}) \odot {\mathbf{I}}.
\end{equation}

With the derived $\nabla {f_2}({\mathbf{\Theta }})$, the next step is to project it onto the tangent
space to obtain the Riemannian gradient and update $\mathbf{\Theta}$ with a proper step size determined by the well-known Armijo backtracking algorithm. Finally, the retraction operation is applied to
make the result satisfy the constant modulus constraint \cite{Yu:2016}. It is worth noting that MO can guarantee the convergence of the objective function to a stationary point where the gradient is zero, according to Theorem 4.3.1 in \cite{Absil:2009}. 

\subsubsection{MM-IRS Algorithm}\label{subsubsec:MM}
Recently, FP has received considerable attention in communication systems \cite{Shen:2019}, especially for solving problems involving ratio terms, as it can reformulate them into a more amicable form to facilitate the solution. In this part, we first translate the problem of (\ref{prb:irs}) into a more tractable form and then propose a low complexity algorithm via MM. 
Utilizing the Woodbury matrix identity, i.e., ${({\mathbf{X + YZ}})^{ - 1}} = {{\mathbf{X}}^{ - 1}} - {{\mathbf{X}}^{ - 1}}{\mathbf{Y}}({\mathbf{I}} + {\mathbf{Z}}{{\mathbf{X}}^{ - 1}}{\mathbf{Y}}){\mathbf{Z}}{{\mathbf{X}}^{ - 1}}$, and ignoring the constant items, the problem of (\ref{prb:irs}) can be rewritten as
 \begin{equation}\label{prb:irs_wood}
\begin{array}{cl}
\displaystyle{\minimize_{{\mathbf{\Theta }}}} & {f_3}({\mathbf{\Theta }}) \\
{\text{s}}{\text{.t}}{\text{.}}& {\theta _i} \in \left[ {0,2\pi } \right],\quad \forall i = 1, \cdots ,N,
\end{array}
\end{equation}
where 
\begin{equation}\label{eqn:f_3}
    {f_3}({\mathbf{\Theta }}) =   -{\text{tr(}}{\mathbf{\Gamma }}{{\mathbf{W}}^H}{\mathbf{H}}{({\mathbf{I}} + {\varphi ^{ - 1}}{{\mathbf{H}}^H}{\mathbf{W\Gamma }}{{\mathbf{W}}^H}{\mathbf{H}})^{ - 1}}{{\mathbf{H}}^H}{\mathbf{W\Gamma }}).
\end{equation}

Although \eqref{eqn:f_3} involves a matrix inverse, it can be equivalently translated into a more tractable form by applying a classical matrix-ratio problem \cite{Shen:2019} with an auxiliary variable ${\mathbf{\Phi }}$. That is,
\begin{equation}\label{prb:irs_FP}
\begin{array}{cl}
\displaystyle{\minimize_{{\mathbf{\Theta }},{\mathbf{\Phi }}}} & {f_4}({\mathbf{\Theta }},{\mathbf{\Phi }}) \\
{\text{s}}{\text{.t}}{\text{.}}&  {\theta _i} \in \left[ {0,2\pi } \right],\quad \forall i = 1, \cdots ,N,
\end{array}
\end{equation}
where
\begin{equation}
    {f_4}({\mathbf{\Theta }},{\mathbf{\Phi }}) = \operatorname{tr} ({{\mathbf{\Phi }}^H}({\mathbf{I}} + {\varphi ^{ - 1}}{{\mathbf{H}}^H}{\mathbf{W\Gamma }}{{\mathbf{W}}^H}{\mathbf{H}}){\mathbf{\Phi }} - 2{\text{Re}}\{ {\mathbf{\Gamma }}{{\mathbf{W}}^H}{\mathbf{H\Phi }}\} )
\end{equation}

In particular, the equivalence between (\ref{prb:irs_wood}) and (\ref{prb:irs_FP}) can be verified by substituting the following optimal ${\mathbf{\Phi }}$ into (\ref{prb:irs_FP})
\begin{equation}
\label{equ:phi}
    {\mathbf{\Phi }} = {({\mathbf{I}} + {\varphi ^{ - 1}}{{\mathbf{H}}^H}{\mathbf{W\Gamma }}{{\mathbf{W}}^H}{\mathbf{H}})^{ - 1}}{{\mathbf{H}}^H}{\mathbf{W\Gamma }}.
\end{equation}

By fixing ${\mathbf{\Phi }}$ and ignoring the constant items in \eqref{prb:irs_FP}, the new problem of optimizing $\mathbf{\Theta}$ can be stated as
 \begin{equation}\label{prb:irs_FP_v1}
\begin{array}{cl}
\displaystyle{\maximize_{{\mathbf{\Theta }}}} & {f_5}({\mathbf{\Theta }}) = \operatorname{tr} ({\mathbf{\Theta }}{{\mathbf{G}}^H} + {{\mathbf{\Theta }}^H}{\mathbf{G}} - {{\mathbf{\Theta }}^H}{\mathbf{D\Theta E}})\\
{\text{s}}{\text{.t}}{\text{.}} & {\theta _i} \in \left[ {0,2\pi } \right],\quad \forall i = 1, \cdots ,N.
\end{array}
\end{equation}
where ${\mathbf{C}} = {\mathbf{H}}_{\text{r}}^H{\mathbf{W\Gamma }}{{\mathbf{\Phi }}^H}{{\mathbf{G}}^H}$, ${\mathbf{D}} = {\varphi ^{ - 1}}{\mathbf{H}}_{\text{r}}^H{\mathbf{W\Gamma }}{{\mathbf{W}}^H}{{\mathbf{H}}_{\text{r}}}$, ${\mathbf{E}} = {\mathbf{G\Phi }}{{\mathbf{\Phi }}^H}{{\mathbf{G}}^H}$, ${\mathbf{F}} = {\varphi ^{ - 1}}{\mathbf{H}}_{\text{r}}^H{\mathbf{W\Gamma }}{{\mathbf{W}}^H}{{\mathbf{H}}_{\text{d}}}{\mathbf{\Phi }}{{\mathbf{\Phi }}^H}{{\mathbf{G}}^H}$, and ${\mathbf{G}} = {\mathbf{C}} - {\mathbf{F}}$.



Using the lemma of matrix identity in \cite{Zhangxianda} that $\operatorname{tr} ({{\mathbf{U}}^H}{\mathbf{Z}}) = {{\mathbf{u}}^H}{\mathbf{z}}$ and $\operatorname{tr} ({{\mathbf{U}}^H}{\mathbf{XUY}}) = {{\mathbf{u}}^H}({\mathbf{X}} \odot {{\mathbf{Y}}^T}){\mathbf{u}}$, where ${\mathbf{u}}$ is an arbitrary vector, ${\mathbf{X}}$, ${\mathbf{Y}}$, and ${\mathbf{Z}}$ are three arbitrary matrices with satisfied sizes for matrix multiplication, ${\mathbf{U}} = {\text{diag}}({\mathbf{u}})$, and ${\mathbf{z}}$ is a column vector consisting of the diagonal elements of $\mathbf{Z}$, the problem in (\ref{prb:irs_FP_v1}) can be rewritten as 
\begin{equation}\label{prb:irs_FP_v2}
\begin{array}{cl}
\displaystyle{\maximize_{{\boldsymbol{\theta}}}} & {f_6}({\boldsymbol{\theta }}) = 2\mathrm{Re} \{ {{\mathbf{g}}^H}{\boldsymbol{\theta }}\}  - {{\boldsymbol{\theta }}^H}{\mathbf{J}\boldsymbol{\theta }}\\
{\text{s}}{\text{.t}}{\text{.}} & {\theta _i} \in \left[ {0,2\pi } \right],\quad \forall i = 1, \cdots ,N,
\end{array}
\end{equation}
where ${\boldsymbol{\theta }} = {\left[ {{e^{j{\theta _1}}}, \cdots ,{e^{j{\theta _N}}}} \right]^T}$, ${\mathbf{g}} = {\text{diag}}({\mathbf{G}})$, and ${\mathbf{J}} = {\mathbf{D}} \odot {{\mathbf{E}}^T}$. 

Although with FP, the objective function in \eqref{prb:irs_FP_v2} has a more concise form than that in \eqref{prb:irs_wood}, it is still difficult to solve because of the constant modulus constraint. However, such constraint can be effectively handled by MM. According to \cite{Sun:2017}, the crucial step is to find a suitable surrogate function for the objective with which MM can find a locally optimal solution of the original objective function. The surrogate function of the objective in (\ref{prb:irs_FP_v2}) can be obtained with the following inequality
\begin{equation}\label{eqn:IRS-MM}
{f_6}(\boldsymbol{\theta}) \geq 2 \mathrm{Re}\{ \mathbf{g}^H\boldsymbol{\theta} \}  - 2\mathrm{Re}\{ \boldsymbol{\theta}^H(\mathbf{J}-\mathbf{L}) \tilde{\boldsymbol{\theta}} - {\boldsymbol{\theta}^H}\mathbf{L}\boldsymbol{\theta} - {\tilde{\boldsymbol{\theta}}^H}(\mathbf{L}-\mathbf{J}){\tilde{\boldsymbol{\theta}}},
\end{equation}
where $\tilde{\boldsymbol{\theta}}$ is a feasible point and $\mathbf{L} \succeq {\mathbf{J}}$ is an auxiliary constant matrix. The equality of \eqref{eqn:IRS-MM} holds when $\boldsymbol{\theta} =\tilde{\boldsymbol{\theta}}$. Furthermore, by setting ${\mathbf{L}} = {\lambda _{\max }}({{\mathbf{J}}}){\mathbf{I}}$ and defining ${\mathbf{q}} = ({\mathbf{L}} - {{\mathbf{J}}})\tilde{\boldsymbol{\theta}} + {\mathbf{g}}$, (\ref{eqn:IRS-MM}) can be rewritten as
\begin{equation}
\label{eqn:IRS-MM_v2}
    {f_6}({\boldsymbol{\theta}}) \geq 2\mathrm{Re} {\{ }{{\boldsymbol{\theta}}^H}{\mathbf{q}}{\} } - 2N{\lambda _{\max }}({{\mathbf{J}}}) + {\tilde{\boldsymbol{\theta }}^H}{\mathbf{J}\tilde{\boldsymbol{\theta}}}.
\end{equation}



Thus, the problem of  (\ref{prb:irs_FP_v2}) can be lowerbounded as 
\begin{equation}\label{prb:irs_FP_v3}
\begin{array}{cl}
\displaystyle{\maximize_{{\boldsymbol{\theta}}}} & {f_7}({\boldsymbol{\theta }}) = 2\mathrm{Re} {\{ }{{\boldsymbol{\theta}}^H}{\mathbf{q}}{\} } - 2N{\lambda _{\max }}({{\mathbf{J}}}) + {\tilde{\boldsymbol{\theta }}^H}{\mathbf{J}\tilde{\boldsymbol{\theta}}}\\
{\text{s}}{\text{.t}}{\text{.}} & {\theta _i} \in \left[ {0,2\pi } \right],\quad \forall i = 1, \cdots ,N.
\end{array}
\end{equation}

By then dropping the last two terms of the objective function in (\ref{prb:irs_FP_v3}) which are irrelevant to  ${\boldsymbol{\theta }}$, the problem in (\ref{prb:irs_FP_v3}) has an optimal solution with a closed expression as follows
\begin{equation}
\label{eqn:theta}
    {\boldsymbol{\theta }}_{\mathrm{opt}} = [ {{e^{j\angle {\mathbf{q}}(1)}}, \cdots ,{e^{j\angle {\mathbf{q}}(N)}}} ]^T.
\end{equation}

It is worth noting that the above MM-IRS algorithm can also guarantee the convergence to a stationary point \cite{Sun:2017}. That is, the solution of \eqref{prb:irs_FP_v3} makes the objective function of \eqref{prb:irs_FP_v2} with a fixed ${\mathbf{\Phi}}$ non-decreasing. Furthermore, as the optimal ${\mathbf{\Phi}}$ in \eqref{equ:phi} establishes the equivalence between  \eqref{prb:irs_wood} and \eqref{prb:irs_FP}, the objective function of (\ref{prb:irs_wood}) is also non-decreasing. The detailed proof of the convergence is provided in the next subsection. The MM-IRS algorithm is summarized in Algorithm \ref{alg:MM}.
 
\begin{algorithm}[t]
\caption{The MM-IRS Algorithm}
\label{alg:MM}
	\begin{algorithmic}[1]
			\STATE Set $i=0$ and initialize ${{\mathbf{\Theta }}^{(i)}}$ randomly;
			\STATE Compute ${\mathbf{\Phi }}$ according to \eqref{equ:phi}; 
\REPEAT	
	\STATE  Compute ${{\mathbf{q}}^{(i)}}$ according to (\ref{prb:irs_FP_v3});
	\STATE  Update ${{\mathbf{\Theta }}^{(i)}}$
	according to (\ref{eqn:theta}); 
    \STATE  $i\leftarrow i+1$;
\UNTIL $(f_7^{(i+1)} - f_7^{(i)})\le \epsilon$; 	\end{algorithmic}
\end{algorithm}

\subsection{Alternating Optimization and Proof of Convergence}\label{subsec:AO}
By alternatively using the transceiver beamforming optimization in Section \ref{subsec:TR-BF} and the MO-IRS or the MM-IRS algorithm in Section \ref{subsec:IRS} for the IRS phase shifts optimization, the MW-MSE minimization problem in \eqref{prb:wmmse} can be finally solved. The whole AO algorithm is summarized in Algorithm \ref{alg:AO}. As shown in Section \ref{subsec:MW-MSE-idea} that there is equivalence between the WSR maximization problem and the MW-MSE minimization one, we now show that the iterative AO algorithm in Algorithm \ref{alg:AO} can guarantee the convergence of the WSR to a locally optimal point. 



\begin{algorithm}[t]
	\caption{AO for MW-MSE minimization}
	\label{alg:AO}
	\begin{algorithmic}[1]
			\STATE Set $t=0$, ${\beta ^{(t)}} = 1$, and initialize ${{\mathbf{F}}^{(t)}}$ and ${{\mathbf{\Theta }}^{(t)}}$ randomly;
\REPEAT	
	\STATE  Compute ${\mathbf{W}}_k^{(t)}$ according to \eqref{beamforming-receiver};
	\STATE  Compute ${\mathbf{\Gamma }}_k^{(t)}$
	according to \eqref{weighted-matrix}; 
    \STATE  Compute ${{\mathbf{\Theta }}^{(t)}}$
	via the MO-IRS or the MM-IRS algorithm;
	\STATE  Compute ${{{\mathbf{\tilde F}}}^{(t)}}$ and ${\beta ^{(t)}}$ according to \eqref{eqn:F}; 
    \STATE  $t\leftarrow t+1$;

\UNTIL $(f_1^{(t)} - f_1^{(t + 1)})\le \epsilon$; 
                \STATE ${\mathbf{F}} = \beta {\mathbf{\tilde F}}$;
	\end{algorithmic}
\end{algorithm}

\begin{proposition}\label{rate-pro}
The proposed AO algorithm in Algorithm \ref{alg:AO} is guaranteed to converge and the WSR is monotonically non-decreasing after each iteration.
\end{proposition}

\textit{Proof}: Please refer to Appendix A.$\hfill\blacksquare$ 

\subsection{Complexity Analysis}\label{subsec:complexity}
In this subsection, we analyze the computational complexity of the proposed AO algorithm in terms of the required number of complex multiplications. We specifically focus on the complexity of the MO-IRS and the MM-IRS algorithms as all variables out of $\mathbf{\Theta}$ are updated based on the closed-form solutions in the AO algorithm. 

\subsubsection{MO-IRS Algorithm}
The complexity of calculating of the Euclidean conjugate gradient in (\ref{eqn:gra-theta}), line search, orthogonal projection, and retraction operations are $2{N^2}K{N_{\text{r}}} + {N^2}{N_{\text{t}}} + 2N{(K{N_{\text{r}}})^2} + 3NK{N_{\text{r}}}{N_{\text{t}}} + \mathcal{O}({K^3}N_{\text{r}}^3)$, $2{(K{N_{\text{r}}})^2}{N_{\text{t}}} + 2\mathcal{O}({K^3}N_{\text{r}}^3)$, ${N^2}$, and ${N}$, respectively, where $\mathcal{O}({K^3}N_{\text{r}}^3)$ represents the complexity of the inversion of a $K{N_{\text{r}}} \times K{N_{\text{r}}}$ matrix. Denoting the number of the conjugate gradient descent iterations and the number of the iterations in the line search as ${N_{{\text{ite,1}}}}$ and ${N_{{\text{ite,2}}}}$ respectively, the overall complexity of the MO-IRS algorithm is ${N_{{\text{ite,1}}}}(2{N^2}K{N_{\text{r}}} + {N^2}{N_{\text{t}}} + {N^2} + N + 3NK{N_{\text{r}}}{N_{\text{t}}} + 2N{(K{N_{\text{r}}})^2} + 2{N_{{\text{ite,2}}}}{(K{N_{\text{r}}})^2}{N_{\text{t}}} + ({N_{{\text{ite,2}}}} + 1)\mathcal{O}({K^3}N_{\text{r}}^3))$.


\subsubsection{MM-IRS Algorithm}
The complexity is mainly dominated by the calculation of $\mathbf{q}$ in (\ref{prb:irs_FP_v3}), which is $2{N^2}K{N_{\text{r}}} + 2{N^2}{N_{\text{t}}} + 2{N^2} + 2N{(K{N_{\text{r}}})^2} + 2NK{N_{\text{r}}}{N_{\text{t}}}$. Denoting the number of iterations as ${N_{{\text{ite}}}}$, the overall complexity of the MM-IRS algorithm is ${N_{{\text{ite}}}}(2{N^2}K{N_{\text{r}}} + 2{N^2}{N_{\text{t}}} + 2{N^2} + 2N{(K{N_{\text{r}}})^2} + 2NK{N_{\text{r}}}{N_{\text{t}}})$. 

Comparing the above analysis of the two IRS optimization algorithms, we can see that as the MM-IRS algorithm results in a closed-form solution when updating $\boldsymbol{\theta}$, it has much lower complexity than the MO-IRS algorithm, especially when the number of  $K$ or $N_{\text{r}}$ is relatively large.

\section{Simulation Results}\label{sec:simulation}
Considering an IRS-aided multiuser MIMO system in a two-dimensional area, where the BS with ${N_{\text{t}}} = 8$ is located at (0m, 0m), and the IRS is located at (250m, 0m). Assume that there are four users uniformly and randomly located in a circle centered at (250m, 30m) with radius 10m and each user has two receive antennas, i.e., ${N_{\text{r}}} = 2$. Similar to \cite{Guo:2020}, we assume that the channels from the BS to all the users follow the Rayleigh fading, and the channel from the BS to the IRS and those from the IRS to the users follow the Rician fading. The Rician factor, the transmission bandwidth, and the noise power spectral density are set to $10\mathrm{dB}$, $180\mathrm{kHz}$, and $-170\mathrm{dBm/Hz}$, respectively. The normalized priority of each user is  chosen inversely proportional to the path-loss component of ${{\mathbf{H}}_{{\text{d}},k}}$. For comparison, we apply the classical WMMSE algorithm in \cite{Christensen:2008} to obtain a benchmark performance for the same system without the aid of IRS (labeled as `No IRS'). Unless otherwise specified, all algorithms adopt the zero-forcing transmit beamforming as the initialization of $\mathbf{F}$, and the thresholds for stopping the iterations of both Algorithm \ref{alg:MM} and Algorithm \ref{alg:AO} are set to $\epsilon = {10^{ - 4}}$. All simulation results are averaged over 100 randomly generated channels.

\begin{figure}[t]
\centering
\begin{minipage}[t]{0.22\textwidth}
\centering
\includegraphics[height=4cm,width=4.35cm]{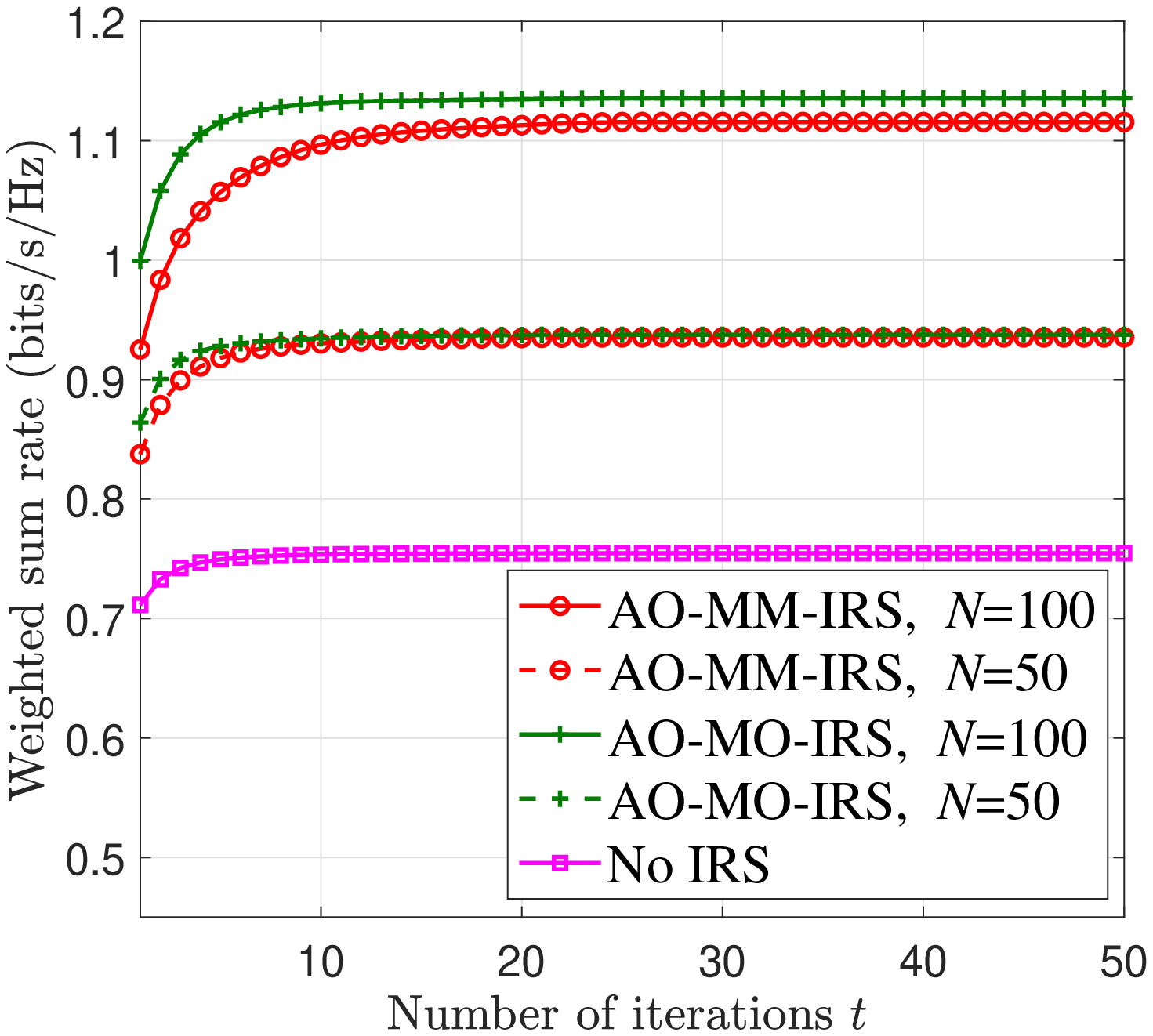}
\caption{Convergence performance of the proposed AO-MO-IRS and AO-MM-IRS algorithms.}
\label{it_sim}
\end{minipage}
\begin{minipage}[t]{0.22\textwidth}
\centering
\includegraphics[height=4cm,width=4.35cm]{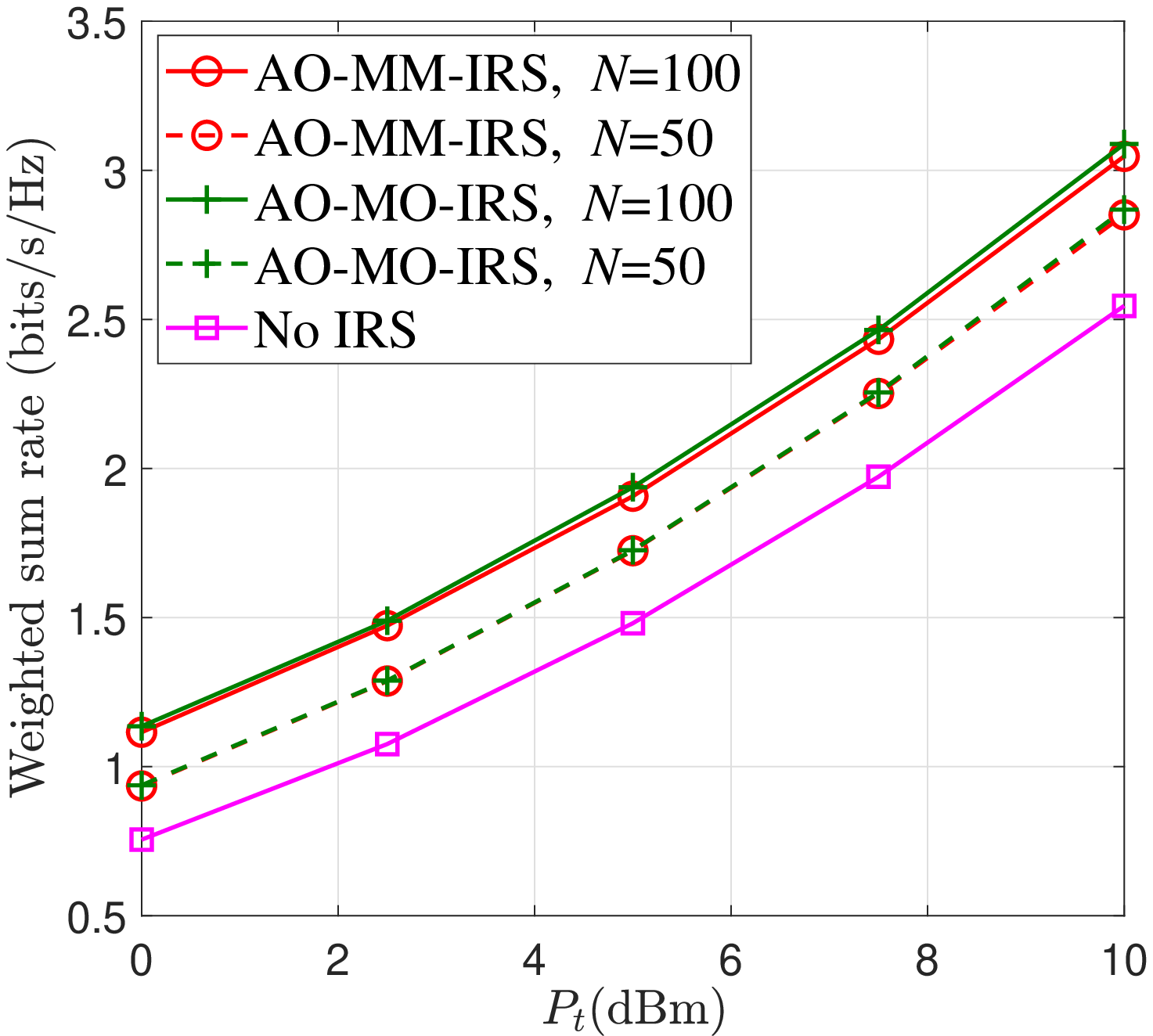}
\caption{WSR v.s. $P_{\text{t}}$ for the proposed AO-MO-IRS and AO-MM-IRS algorithms.}
\label{rate_sim}
\end{minipage}
\end{figure}

To evaluate the convergence of the two proposed algorithms (labeled as `AO-MO-IRS' and `AO-MM-IRS'), we show in Fig. \ref{it_sim} the resulting WSR as a function of the number of alternating iterations when ${P_{\text{t}}}=0\mathrm{dBm}$. The results clearly verify the convergence proof in Appendix A. We can also see that as the number of IRS elements increases, more iterations are needed but higher WSR is achieved.

Fig. \ref{rate_sim} further illustrates the WSR as a function of the transmit power $P_{\text{t}}$ for the proposed algorithms. We can see from this figure that with the joint active and passive beamforming, the WSR can be greatly improved when compared to the system without IRS assistance, and such performance improvement gets higher with a larger number of IRS elements. We can also see that the performance gap between the two proposed algorithms is very limited. However, as shown in Section \ref{subsec:complexity}, the AO-MM-IRS algorithm has much lower computational complexity. Thus, the AO-MM-IRS algorithm has more potential for application in practical systems.



\section{Conclusion}\label{sec:conclusion}
We have investigated the joint active and passive beamforming optimization to maximize the WSR for IRS-aided multiuser MIMO systems via an MW-MSE minimization equivalence. We have shown that the active beamforming subproblem has closed-form solutions and proposed two iterative algorithms for the passive IRS beamforming subproblem with guaranteed convergence by using MO and MM. Simulation results have shown that both the proposed AO-MO-IRS and AO-MM-IRS algorithms significantly outperform the conventional algorithm in the scenario without IRS assistance. 


\section*{Appendix A}\label{APA}
It is worth noting that for the IRS optimization part in the AO, as the MO-IRS algorithm directly optimizes the objective function and is itself a gradient descend algorithm, the convergence proof is relatively easier than that for the MM-IRS algorithm, where the objective function is changed by FP. Thus, we take the AO with the MM-IRS algorithm (AO-MM-IRS) as an example to prove the convergence on the WSR performance.

As shown in Algorithm \ref{alg:AO}, a superscript $t$ is introduced for the optimization variables to denote the iteration index. Defining ${{\mathbf{\Phi }}^{(t+1)}}$ and ${\mathbf{\tilde{\Phi}}}^{(t+1)}$ as the result by respectively substituting  $({{\mathbf{W}}^{(t+1)}},{{\mathbf{\Gamma }}^{(t+1)}},{{\mathbf{\Theta }}^{(t)}},{{{\mathbf{\tilde F}}}^{(t)}},{\beta ^{(t)}})$ and  $({{\mathbf{W}}^{(t+1)}},{{\mathbf{\Gamma }}^{(t+1)}},{{\mathbf{\Theta }}^{(t + 1)}},{{{\mathbf{\tilde F}}}^{(t)}},{\beta ^{(t)}})$ into \eqref{equ:phi}, we have 

\begin{equation*}
    \begin{aligned}
    &{f_1}({{\mathbf{W}}^{(t)}},{{\mathbf{\Gamma }}^{(t)}},{{\mathbf{\Theta }}^{(t)}},{{{\mathbf{\tilde F}}}^{(t)}},{\beta ^{(t)}})\\ \stackrel{(a)}\geq& {f_1}({{\mathbf{W}}^{(t+1)}},{{\mathbf{\Gamma }}^{(t)}},{{\mathbf{\Theta }}^{(t)}},{{{\mathbf{\tilde F}}}^{(t)}},{\beta ^{(t)}})\\ \stackrel{(b)}\geq& {f_1}({{\mathbf{W}}^{(t+1)}},{{\mathbf{\Gamma }}^{(t +1)}},{{\mathbf{\Theta}}^{(t)}},{{{\mathbf{\tilde F}}}^{(t)}},{\beta ^{(t)}})\\ \stackrel{(c)}{=}& {f_4}({{\mathbf{W}}^{(t+1)}},{{\mathbf{\Gamma }}^{(t+1)}},{{{\mathbf{\Phi }}}^{(t+1)}},{{\mathbf{\Theta }}^{(t)}},{{{\mathbf{\tilde F}}}^{(t)}},{\beta ^{(t)}})\\ \stackrel{(d)}\geq& {f_4}({{\mathbf{W}}^{(t+1)}},{{\mathbf{\Gamma }}^{(t+1)}},{{\mathbf{\Phi }}^{(t+1)}},{{\mathbf{\Theta}}^{(t+ 1)}},{{{\mathbf{\tilde F}}}^{(t)}},{\beta ^{(t )}})\\ \stackrel{(e)}\geq& {f_4}({{\mathbf{W}}^{(t+1)}},{{\mathbf{\Gamma }}^{(t+1)}},{{{\mathbf{\tilde \Phi }}}^{(t+1)}},{{\mathbf{\Theta }}^{(t+1)}},{{{\mathbf{\tilde F}}}^{(t)}},{\beta ^{(t )}})\\ \stackrel{(f)}{=}& {f_1}({{\mathbf{W}}^{(t+1)}},{{\mathbf{\Gamma }}^{(t+1)}},{{\mathbf{\Theta }}^{(t+1)}},{{{\mathbf{\tilde F}}}^{(t)}},{\beta ^{(t)}})\\  \stackrel{(g)}\geq& {f_1}({{\mathbf{W}}^{(t+1)}},{{\mathbf{\Gamma }}^{(t+1)}},{{\mathbf{\Theta }}^{(t+1)}},{{{\mathbf{\tilde F}}}^{(t+1)}},{\beta ^{(t+1)}}),
    \end{aligned}
\end{equation*}
where ($a$) and ($b$) follow from the fact that the closed-form solutions satisfying the KKT conditions always ensure the non-increase of the objective function, ($c$) holds because the ${\mathbf{\Phi}}$ computed from (\ref{equ:phi}) establishes the equivalence between  \eqref{prb:irs_wood} and \eqref{prb:irs_FP}, ($d$) follows from the fact that the convergent update of ${\mathbf{\Theta }}$ via MM ensures the non-increase of $f_4$ when other variables are fixed \cite{Sun:2017}, ($e$) follows from the fact that the update of ${\mathbf{\Phi}}$ minimizes $f_{4}$ when other variables are fixed, ($f$) holds with the same reason as that of ($c$), and finally ($g$) follows with the same reason as that of ($a$) and ($b$).

Next, by substituting the closed-form solution of ${{\mathbf{W}}_k}$ in (\ref{beamforming-receiver}) into (\ref{MSE_matrix}), the MSE matrix can be expressed as ${{\mathbf{E}}_k} = {({\mathbf{I}} + {\mathbf{F}}_k^H{\mathbf{H}}_k^H{\mathbf{\Lambda}}_k^{ - 1}{{\mathbf{H}}_k}{{\mathbf{F}}_k})^{ - 1}}$. Then, according to (\ref{weighted-matrix}), the function value of $f_1$ can be expressed as
\begin{equation}
\label{eqva-rate}
    \begin{aligned}
{f_1}& = \sum\limits_{k = 1}^K {{\omega _k}{N_{\text{r}}} - } \sum\limits_{k = 1}^K {{\omega _k}\log \left| {{\mathbf{I}} + {\mathbf{F}}_k^H{\mathbf{H}}_k^H{\mathbf{\Lambda}}_k^{ - 1}{{\mathbf{H}}_k}{{\mathbf{F}}_k}} \right|} \\ &= \sum\limits_{k = 1}^K {{\omega _k}{N_{\text{r}}} - } R.
    \end{aligned}
\end{equation}

According to the relationship between $f_1$ and the WSR $R$ in  (\ref{eqva-rate}), and the above proof of the convergence of $f_1$, we can see that the WSR is monotonically non-decreasing in the iterations. The proof is thus completed. $\hfill\blacksquare$

\bibliography{ref}
\bibliographystyle{IEEEtran}
\end{document}